\documentclass[11pt,psfig]{article}

\usepackage{graphicx}
\input epsf.tex

\usepackage{epsfig,fancybox}
\usepackage{amsmath}
\usepackage{amssymb}
\usepackage{amsfonts}
\usepackage{graphicx}
\usepackage{dsfont}
\usepackage{caption}
\usepackage{subcaption}
\usepackage{color}

\newtheorem{thm}{Theorem}[section]

\newtheorem{lem}{Lemma}[section]

\newcommand{\lemref}[1]{Lemma~{\rm \ref{#1}}}

\newcounter{neweqn}

\newcommand{\beq}[1]{\begin{equation} \refstepcounter{neweqn} \label{#1}}
\newcommand{\eeq}{\end{equation}}
\newcommand{\bed}{\begin{displaymath}}
\newcommand{\eed}{\end{displaymath}}
\newcommand{\bedd}{\bed\begin{array}{l}}
\newcommand{\eedd}{\end{array}\eed}
\newcommand{\nd}{\noindent}

\newcommand{\rf}[1]{(\ref{#1})}
\newcommand{\disp}{\displaystyle}
\newcommand{\A}{{\mathcal A}}

\newcommand{\bfS}{{\mathbf S}}

\newcommand{\Bb}{\beta_{\rm b}}
\newcommand{\Bs}{\beta_{\rm s}}

\newcommand{\pdev}[2]{\frac{\partial #1}{\partial #2}}

\newcommand{\rr}{{\hbox{{\rm I}{\kern -0.22em}{\rm R}}}}
\newcommand{\bdd}{\hspace*{-0.08in}{\bf.}\hspace*{0.05in}}

\newcommand{\al}{\alpha}
\newcommand{\la}{\lambda}

\def\({\left(}
\def\){\right)}

\def \fA {{\mathcal A}}
\def \F {{\mathcal F}}
\def \L {{\mathcal L}}
\def \cal {\mathcal}
\def \M {{\mathcal M}}

\begin{document}

\title{Pairs Trading: An Optimal Selling Rule with Constraints}

\author{R. Liu\thanks{School of Mathematics and Statistics, University of Sydney, Sydney, NSW 2006, Australia, E-mail: ruyi.liu@sydney.edu.au}
  \and
  J. Tie\thanks{Department of Mathematics, University of Georgia, Athens, GA 30602, USA, E-mail: jtie@uga.edu}
  \and Z. Wu\thanks{School of Mathematics, Shandong University,
    Jinan {\rm 250100}, P.R. China, E-mail: wuzhen@sdu.edu.cn. }
  \and
  Q. Zhang\thanks{Department of Mathematics, University of Georgia, Athens, GA 30602, USA, E-mail: qz@uga.edu}
}

\maketitle
    
\begin{abstract}
  The focus of this paper is on identifying the most effective selling strategy for pairs trading of stocks. In pairs trading, a long position is held in one stock while a short position is held in another. The goal is to determine the optimal time to sell the long position and repurchase the short position in order to close the pairs position. The paper presents an optimal pairs-trading selling rule with trading constraints. In particular, the underlying stock prices evolve according to a two dimensional geometric Brownian motion and the trading permission process is given in terms of a two-state $\{$trading allowed,
  trading not allowed$\}$ Markov chain. It is shown that the optimal policy can be determined by a threshold curve which is obtained by solving the associated HJB equations (quasi-variational inequalities). A closed form solution is obtained. A verification theorem is provided. Numerical experiments are also reported to demonstrate the optimal policies and value functions.
  

\end{abstract}

\section{Introduction} 
Pairs trading involves the simultaneous trading of two stocks as a pair. Typically, a pairs position involves taking a long position in one stock and a short position in the other. The focus of this paper is on determining the optimal timing for closing an existing pairs position. Specifically, we aim to identify the most effective strategy for folding the position and exiting the market.

Pairs trading is closely related to the timing of the optimal investments
studied in McDonald and Siegel \cite{McDonaldS}.
In particular, they considered the optimal
timing of investment in an irreversible project. 
Two main variables in their model are the value of the project
and the cost of investing.  They demonstrated one should defer the investment until the present value of the 
benefits from the project exceed the investment cost by a certain margin. They transfer the control problem in selecting the optimal trading time to the one of determining the optimal value in space.
Further studies along this line were carried out by
Hu and \O ksendal \cite{HuO} to specify precise optimality conditions
and to provide new proof of the following variational inequalities among others.
Their results can be easily interpreted in terms of the pairs-trade selling rule
when treating the project value as the long position and investment cost as the short
position.

In this paper, we extend these results to incorporate markets with Markov trading
constraints. In particular, a sequence of trading windows is imposed. One
can only buy/sell stocks when the windows are open.
We focus on a simple and easily implementable strategy, and its optimality and 
the sufficient conditions for a closed-form solution.

Mathematical trading rules have been studied extensively in the literature, and various approaches have been proposed to determine optimal trading strategies.
Zhang \cite{Zhang-trading} considers a selling rule that involves two threshold levels, a target price, and a stop-loss limit, and aims to find the optimal threshold levels that maximize the expected profit. To achieve this, Zhang solves a set of two-point boundary value problems, which are a type of differential equation with conditions specified at two endpoints. The resulting optimal threshold levels can then be used to construct a profitable trading strategy.
In \cite{GuoZ}, Guo and Zhang study the optimal selling rule under a model with switching Geometric Brownian motion, which is a type of stochastic process commonly used to model asset prices. To determine the optimal threshold levels, they use a smooth-fit technique that involves solving a set of algebraic equations. This approach allows them to obtain explicit expressions for the threshold levels, which can be used to design a profitable trading strategy.
Dai et al. \cite{DaiZZ} proposed a trend-following rule based on a conditional probability indicator. They showed that the optimal trading rule can be determined by solving the associated Hamilton-Jacobi-Bellman equations, which are a type of partial differential equation commonly used in finance.
Iwarere and Barmish \cite{IwarereB} developed a similar idea using a confidence interval approach, while Merhi and Zervos \cite{MerhiZ} studied an investment capacity expansion/reduction problem using dynamic programming under a geometric Brownian motion market model.
In the context of mean reversion trading, Zhang and Zhang \cite{ZhangZ}
obtained a buy-low and sell-high policy by characterizing the 'low' and 'high' levels in terms of mean reversion parameters.
Song and Zhang \cite{SongZ} studied pairs trading under a mean reversion model. It is shown that the optimal trading rule can be determined by threshold levels that can be obtained by solving a set of algebraic equations. A set of sufficient conditions are also provided to establish the desired optimality.
Deshpande and Barmish \cite{DeshpandeB} developed a control-theoretic approach to pairs trading that relaxes the requirement for spread functions. They demonstrated that their trading algorithm generates positive expected returns. Other pairs trading methods can be found in Elliott et al. \cite{Elliott}
and Whistler \cite{Whistler}.
More recently, Tie et al. \cite{TieZZ}
proposed an optimal pairs trading rule aimed at maximizing a discounted payoff function by sequentially initiating and closing positions of the pair. Using a dynamic programming approach under a geometric Brownian motion model, they showed that buying and selling decisions can be determined by two threshold curves in closed form. They also proved the optimality of their trading strategy.

Studies of trading rules with trading constraints are important from application point of view.
Market liquidity is one of the causes of limited trading windows.
Poor liquidity restricts one's ability to trade freely.
Such trading conditions have greater impacts on small cap stocks.
It affects even more on pairs trading because both buying and selling
have to executed all together.
Mathematical treatment on trading constraints can be found in
Dupuis and Wang \cite{DupuisW} in which they used a Poisson 
process to capture permissible trading moments. They obtained
a closed form solution under a one-dimensional geometric Brownian motion.
Further studies can be found in Menaldi and Robin \cite{MenaldiR}.
They treated related optimal stopping with a class of general Markov-Feller processes and focused on theoretical characterizations of optimal stopping. 


In this paper, we consider a pairs selling rule under a two-dimensional
geometric Brownian motion with trading constraints.
We model the sequence of trading windows in terms of a two-state
Markov chain. One can only buy/sell stocks when the trading windows
are open. We focus on such constrained optimal stopping.
We generalize the results of
Hu and Oksendal \cite{HuO} by imposing trading constraints.
We show that the optimal selling rule can be determined by a single 
threshold curve. We also establish sufficient conditions that
guarantee the optimality of the selling policy.
In addition, we report our numerical experiments to demonstrate
our results.

This paper is organized as follows.
In \S2, we formulate the pairs trading problem under consideration.
In \S3, we study the associated HJB equations and their solutions.
In \S4, we provide a verification theorem 
that guarantee the optimality of our selling rule.
In \S5, we consider asymptotic properties of the trading switching policies
as the trading constraint jump rates go to infinity.
Numerical examples are given in \S6.
Some concluding remarks are given in \S7. 
We postpone the proof of lemma 3.1 to the appendix.

\section{Problem Formulation}
Our pairs trading involves two stocks: $\bfS^1$ and $\bfS^2$.
Let $\{X^1_t,t\geq0\}$ denote the prices of stock $\bfS^1$ and
$\{X^2_t,t\geq0\}$ that of stock $\bfS^2$.
They satisfy the following stochastic differential equation:
\beq{sys-eqn}
d\(\!\!\begin{array}{c}
X^1_t\\
X^2_t\\
\end{array}\!\!\)
=
\(\!\!\begin{array}{cc}
X^1_t& \\
     &X^2_t\\
\end{array}\!\!\)
\left[
\(\!\!\begin{array}{c}
\mu_1\\
\mu_2\\
\end{array}\!\!\)dt
+
\(\!\!\begin{array}{cc}
\sigma_{11}&\sigma_{12}\\
\sigma_{21}&\sigma_{22}\\
\end{array}\!\!\)
d\(\!\!\begin{array}{c}
W^1_t\\
W^2_t\\
\end{array}\!\!\)
\right],
\eeq
where $\mu_i$, $i=1,2$, are the return rates, $\sigma_{ij}$, $i,j=1,2$,
the volatility constants, 
and $(W^1_t,W^2_t)$ a 2-dimensional
standard Brownian motion.

The liquidity process $\al_t$ is assumed to be a two-state Markov chain with state space $\M=\{0,1\}$.
We impose the following trading constraint: One can only buy/sell stocks
when $\al_t=1$.
Let $Q$ be the generator of $\al_t$ given by
$Q=\left(\begin{array}{cc}
-\la_0 & \la_0\\[-0.05in]
\la_1 & -\la_1\\
\end{array} \right)$, with $\la_0>0$ and $\la_1>0$.
We assume $\al_t$ and $(W^1_t,W^2_t)$ are independent.

In this paper, we consider a pairs selling rule. 
We assume the corresponding pair's position consists of
a one-share long position in stock
$\bfS^1$ and a one-share short position in stock $\bfS^2$.
This condition can be easily relaxed; see Tie et al. \cite{TieZZ} for
details. The problem is to determine an optimal stopping time $\tau$
(subject to trading constraints) to fold
the pairs  position by selling $\bfS^1$ and buying back $\bfS^2$.

Let $K$ denote the transaction cost percentage (e.g., slippage and/or
commission) associated with stock transactions.
For example, the proceeds to close the pairs position at $t$ is
$(1-K)X^1_{t}-(1+K)X^2_{t}$.
For ease of notation, let $\Bb=1+K$ and $\Bs =1-K$.

Let $\F_t=\sigma\{(X^1_r,X^2_r,\al_r):\ r\leq t\}$.
We consider admissible stopping times ${\cal S}=\{\tau:\ {\cal F}_t
\mbox{ stopping times such that }\tau<\infty \mbox{ only when }
\al_\tau=1\}$.

Given the initial state $(x_1,x_2)$, $\al=0,1$, and the admissible
selling time $\tau$,
the corresponding reward function
\beq{reward-fn}
J(x_1,x_2,\al,\tau)=
E\big[e^{-\rho\tau}(\Bs X^1_{\tau}-\Bb X^2_{\tau})\big],
\eeq
where $\rho>0$ is a given discount factor.

The problem is to find
an admissible stopping time $\tau$ to maximize $J$.
Let $V_\al(x_1,x_2)$ denote the corresponding value function:
\beq{value-fn}
V_\al(x_1,x_2)=\sup_{\tau}J(x_1,x_2,\al,\tau).
\eeq

We impose the following conditions throughout this paper.

{\bf (A1)} $\rho>\mu_1$ and $\rho>\mu_2$.

Under these conditions, we have the lower and upper bounds for $V$:
\beq{bounds}
-\Bb  x_2\leq V_\al(x_1,x_2)\leq \Bs  x_1.
\eeq
Actually, for any $\tau\in{\cal S}$, we have
$-\Bb E\big[e^{-\rho\tau}X^2_{\tau}\big]\leq J(x_1,x_2,\al,\tau)\leq
\Bs E\big[e^{-\rho\tau}X^1_{\tau}\big]$.
The rest follows from Dynkin's formula
\[
E[e^{-\rho\tau} X^j_{\tau}]
=\(x_j+E\int_0^\tau e^{-\rho t}X^j_t(-\rho+\mu_j)dt\) \leq x_j,\mbox{ for }j=1,2.
\]

In addition, in view of the value function definition, we have
\[
V_1(x_1,x_2)\geq J(x_1,x_2,1,\tau)|_{\tau=0}=\Bs x_1-\Bb x_2.
\]

\section{HJB Equations}
In this paper, we follow the dynamic programming approach and
study the associated HJB equations.
Let
\[
\A=\frac{1}{2}\left\{a_{11}x_1^2\frac{\partial^2}{\partial x_1^2}
+2a_{12}x_1x_2\frac{\partial^2}{\partial x_1 \partial x_2}
+a_{22}x_2^2\frac{\partial^2}{\partial x_2^2}\right\}
+\mu_1 x_1\frac{\partial}{\partial x_1}
+\mu_2 x_2\frac{\partial}{\partial x_2},
\]
where
$a_{11}=\sigma^2_{11}+\sigma^2_{12},\
a_{12}=\sigma_{11}\sigma_{21}+\sigma_{12}\sigma_{22},\mbox{ and }
a_{22}=\sigma^2_{21}+\sigma^2_{22}$.
The associated HJB equations have the form:
For $x_1,x_2>0$,
\beq{HJB}
\left\{\begin{array}{l}
[(\rho+\lambda_0)-\A] v_0(x_1,x_2) =\lambda_0 v_1(x_1,x_2),\\

\min\Big\{[\rho+\lambda_1)-\A] v_1(x_1,x_2)-\lambda_1v_0(x_1,x_2),\
v_1(x_1,x_2)-\Bs  x_1+\Bb  x_2\Big\}=0.\\
\end{array}\right.
\eeq

To solve the above HJB equations, we first convert them into single variable
equations. Let $y=x_2/x_1$ and $v_i(x_1,x_2)=x_1w_i(x_2/x_1)$,
for some function $w_i(y)$ and $i=0,1$.
Then we have by direct calculation that
\[
\begin{array}{l}
\disp
\pdev{v_i}{x_1}=w_i(y)-yw'_i(y),\
\pdev{v_i}{x_2}=w'_i(y),\\
\disp
\pdev{^2v_i}{x_1^2}=\frac{y^2 w''_i(y)}{x_1},\
\pdev{^2v_i}{x^2_2}=\frac{w''_i(y)}{x_1},\mbox{ and }
\pdev{^2v_1}{x_1\partial x_2}=-\frac{yw''_i(y)}{x_1}.
\end{array}
\]
Write $\fA v_i$ in terms of $w_i$ to obtain
\[
\fA v_i=x_1\left\{\frac{1}{2}\left[a_{11}-2a_{12}+a_{22}\right]y^2w''_i(y)+(\mu_2-\mu_1)y w'_i(y)+\mu_1w_i(y)\right\}.
\]
Then, the HJB equations can be given in terms of $y$ and $w_i$ as follows:
\beq{HJB-ode}
\left\{\begin{array}{l}
[(\rho+\lambda_0)-\L] w_0(y) =\lambda_0 w_1(y),\\
\min\Big\{[(\rho+\lambda_1)-\L] w_1(y)-\lambda_1w_0(y),\
w_1(y)-\Bs  +\Bb  y\Big\}=0,\\
\end{array}\right.
\eeq
where 
\[\L [w_i(y)]=\sigma y^2w''_i(y)+(\mu_2-\mu_1)y w'_i(y)+\mu_1w_i(y),
\]
with $\sigma=({a_{11}-2a_{12}+a_{22}})/{2}\geq0$.
We only consider the case when $\sigma\not=0$.
If $\sigma=0$, the problem reduces to a first-order case and
can be treated in a similar way.

\subsection*{Solution of the HJB Equation}

Intuitively, one should close the pairs position when $X^1_t$ is large and
$X^2_t$ is small. Namely, we expect a threshold $k$ so that the pairs position
is to be closed when $y=x_2/x_1\leq k$. 
In view of this, we focus on searching for such $k$ so that
the second equation in \rf{HJB-ode} has the form
$w_1(y)=\Bs-\Bb y$, for $y\in(0,k)$; and 
$[(\rho+\lambda_1)-\L] w_1(y)=\lambda_1w_0(y)$, for $y\in(k,\infty)$,
Since $w_0(y)$ satisfies
\[[(\rho+\lambda_0)-\L] w_0(y) =\lambda_0 w_1(y)\]
for all $y$, we can combine these equations on the interval $(k,\infty)$
and get a system of $w_0(y)$ and $w_1(y)$:
\beq{System-1}
[(\rho+\lambda_1)-\L] w_1(y)=\lambda_1w_0(y)\quad \text{and}\quad
[(\rho+\lambda_0)-\L] w_0(y) =\lambda_0 w_1(y)
\eeq
We can reduce the system into a single equation about $w_1(y)$ (or $w_0(y)$);
\[\{[(\rho+\lambda_0)-\L][(\rho+\lambda_1)-\L]-\lambda_0 \lambda_1\} w_1(y)=0.\]
The above equation can be simplified to
\[[(\L-\rho)(\L-\rho-\lambda_0-\lambda_1)]w_1(y)=0.\]

This equation is a Cauchy-Euler type equation and its solutions are the linear combination of
$y^{\delta}$ with $\delta$ satisfying the polynomial equation:
\beq{delta-eq}
[\sigma \delta(\delta-1)+(\mu_2-\mu_1)\delta+\mu_1-\rho]
[\sigma \delta(\delta-1)+(\mu_2-\mu_1)\delta+\mu_1-\rho-\lambda_0-\lambda_1]=0.
\eeq
Simplify further and we can find $\delta$ explicitly:
\[\left[\delta^2-\left(1+\frac{\mu_1-\mu_2}{\sigma}\right) \delta-\frac{\rho-\mu_1}{\sigma}\right]
\left[\delta^2-\left(1+\frac{\mu_1-\mu_2}{\sigma}\right) \delta-
\frac{\rho+\lambda_0+\lambda_1-\mu_1}{\sigma}\right]=0.\]

There are four real roots $\delta_1$, $\delta_2$, $\delta_3$ and $\delta_4$
(by direct calculation $\delta_1<0, \delta_3<0$ and $\delta_2>1, \delta_4>1$) given by
\beq{deltas}
\begin{array}{l}
\disp
\delta_1=\frac{1}{2}\Biggl(1+\frac{\mu_1-\mu_2}{\sigma}-\sqrt{\(1+\frac{\mu_1-\mu_2}{\sigma}\)^2+\frac{4(\rho-\mu_1)}{\sigma}}\,\Biggr),\\
\disp
\delta_2=\frac{1}{2}\Biggl(1+\frac{\mu_1-\mu_2}{\sigma}+
\sqrt{\(1+\frac{\mu_1-\mu_2}{\sigma}\)^2+\frac{4(\rho-\mu_1)}{\sigma}}\,\Biggr),\\
\disp
\delta_3=\frac{1}{2}\Biggl(1+\frac{\mu_1-\mu_2}{\sigma}-\sqrt{\(1+\frac{\mu_1-\mu_2}{\sigma}\)^2+\frac{4(\rho+\lambda_0+\lambda_1-\mu_1)}{\sigma}}\,\Biggr),\\
\disp
\delta_4=\frac{1}{2}\Biggl(1+\frac{\mu_1-\mu_2}{\sigma}+
\sqrt{\(1+\frac{\mu_1-\mu_2}{\sigma}\)^2+\frac{4(\rho+\lambda_0+\lambda_1-\mu_1)}{\sigma}}\,\Biggr).
\end{array}
\eeq

Using the inequalities in \rf{bounds}, we have $-\Bb y \leq w_i(y)\leq \Bs$.
In view of this, 
the general solution of $w_1(y)$
for $y\in(k,\infty)$ should be of the form:
\beq{w1}
w_1(y)=C_1y^{\delta_1}+C_3 y^{\delta_3},
\eeq
for some constants $C_1$ and $C_3$.
Once we find $w_1(y)$, $w_0(y)$ is given by
\[w_0(y)=\frac{[(\rho+\lambda_1)-\L] w_1(y)}{\lambda_1}=w_1(y)-\frac{1}{\lambda_1} (\L-\rho)w_1(y).\]
Note that $w_1(y)=C_1y^{\delta_1}+C_3y^{\delta_3}$ and 
\[(\L-\rho)(y^{\delta_1})=0\quad\text{and}\quad
(\L-\rho-\lambda_0-\lambda_1)y^{\delta_3}=0.\]
This yields
\begin{align*}
w_0(y)&=w_1(y)-\frac{1}{\lambda_1} (\L-\rho)w_1(y)\\
&=C_1y^{\delta_1}+C_3 y^{\delta_3}-\frac{1}{\lambda_1} (\L-\rho)[C_1y^{\delta_1}+C_3 y^{\delta_3}]\\
&=C_1y^{\delta_1}+C_3 y^{\delta_3}-\frac{1}{\lambda_1}(\L-\rho)[C_3 y^{\delta_3}]\\
&=C_1y^{\delta_1}+C_3 y^{\delta_3}-\frac{\lambda_0+\lambda_1}{\lambda_1} C_3 y^{\delta_3}\\
&=C_1y^{\delta_1}-\frac{\lambda_0}{\lambda_1} C_3 y^{\delta_3}.
\end{align*}
Let $\eta=({\lambda_0}/{\lambda_1})$. Then, we have
\beq{w0w1}
w_0(y)=C_1y^{\delta_1}-\eta C_3 y^{\delta_3}\quad\text{and}\quad
w_1(y)=C_1y^{\delta_1}+C_3 y^{\delta_3}
\eeq
on the interval $(k,\infty)$.

On the interval $(0,k)$, $w_1(y)=\Bs-\Bb  y$ and
\[[(\rho+\lambda_0)-\L] w_0(y) =\lambda_0 w_1(y)=\lambda_0 (\Bs-\Bb y).\]
A particular solution for
\[
(\rho+\lambda_0 -\L )w_0(y)=\lambda_0(\Bs-\Bb y)
\]
can be given by
$w_0(y)=a_0\Bs -a_1 \Bb y$, with
\beq{a1a2}
a_0=\frac{\lambda_0}{\rho+\lambda_0-\mu_1}
\quad \text{and}\quad a_1=\frac{\lambda_0}{\rho+\lambda_0-\mu_2}.
\eeq
To find a general solution to the above non-homogeneous equation,
we only need to add the homogeneous equation $(\rho+\lambda_0-\L)w_0=0$ to the above particular
solution. 
This is also of Cauchy-Euler type and its solution is of the form
$y^{\gamma}$. 
Then $\gamma$ must be the roots of the quadratic equation
\[\sigma \gamma(\gamma-1)+(\mu_2-\mu_1)\gamma-(\rho+\lambda_0-\mu_1)=0.
\]
and are given by 
\beq{gamma1-gamma2}
\begin{split}
\gamma_1&=\frac{1}{2}\left(1+\frac{\mu_1-\mu_2}{\sigma}
-\sqrt{\left(1+\frac{\mu_1-\mu_2}{\sigma}\right)^2+\frac{4(\rho+\lambda_0-\mu_1)}{\sigma}}\right) <0,\\
\gamma_2&=\frac{1}{2}\left(1+\frac{\mu_1-\mu_2}{\sigma}
+\sqrt{\left(1+\frac{\mu_1-\mu_2}{\sigma}\right)^2+\frac{4(\rho+\lambda_0-\mu_1)}{\sigma}}
\right)>1.
\end{split}
\eeq
The general solution for $w_0(y)$ on $(0,k)$ is given by 
\beq{w0inR1}
w_0=C_2 y^{\gamma_2}+a_0\Bs-a_1\Bb y,
\eeq
for some constant $C_2$.
Here, we skipped the term $y^{\gamma_1}$ because the solution
should be bounded at $y=0$. We summarize the computation so far as follows.
  
The solutions of the HJB equations \rf{HJB-ode} have the following form:
\beq{solution}
\begin{array}{l}
w_0(y)=\left\{\begin{array}{ll}
  C_2y^{\gamma_2}+a_0\Bs-a_1\Bb y&\mbox{ if }0<y<k,\\
  C_1y^{\delta_1}-\eta C_3y^{\delta_3}&\mbox{ if }y\geq k,\\
\end{array}\right.\\
w_1(y)=\left\{\begin{array}{ll}
  \Bs-\Bb y&\mbox{ if }0<y<k,\\
  C_1y^{\delta_1}+C_3y^{\delta_3}\hspace*{0.53in}&\mbox{ if }y\geq k.\\
  \end{array}\right.\\
\end{array}
\eeq
In what follows, we use a smooth-fit approach to determine the parameter values
$k$, and $C_j$, $j=1,2,3$.

\subsection*{Smooth-fit conditions.}
Smooth-fit conditions in connection with optimal stopping typically 
require the value functions to be continuously differentiable.
First, the continuous differentiability of $w_1$ at $y=k$ yields
\beq{first-two}
\begin{array}{rl}
\Bs-\Bb k =&C_1k^{\delta_1}+C_3k^{\delta_3},\\
-\Bb k = &C_1\delta_1k^{\delta_1}+C_3\delta_3 k^{\delta_3}.\\
\end{array}
\eeq
Similarly, continuous differentiability of $w_0$ at $y=k$ leads to
\beq{second-two}
\begin{split}
C_2 k^{\gamma_2}+
a_0\Bs-a_1\Bb k=& C_1 k^{\delta_1}-\eta C_3 k^{\delta_3},\\
C_2 \gamma_2 k^{\gamma_2}
-a_1\Bb k=&\delta_1C_1k^{\delta_1}-\eta\delta_3 C_3 k^{\delta_3}.
\end{split}
\eeq
Here we have four equations with four unknown parameters $k$, $C_j$ for $j=1,2,3$.

Let
\[\Phi(k;\delta_1,\delta_3)=\begin{pmatrix}
k^{\delta_1} & k^{\delta_3}\\
\delta_1 k^{\delta_1} & \delta_3 k^{\delta_3}
\end{pmatrix}.\]
Its inverse is given by
\[\Phi^{-1}(k;\delta_1,\delta_3)=\frac{1}{\delta_3-\delta_1} \begin{pmatrix}
\delta_3 k^{-\delta_1} & -k^{-\delta_1}\\
-\delta_1 k^{-\delta_3} &  k^{-\delta_3}
\end{pmatrix}.\]
Using $\Phi$, we rewrite \rf{first-two} and \rf{second-two}:
\[
\Phi(k;\delta_1,\delta_3)\begin{pmatrix}
C_1\\ C_3\end{pmatrix} =\begin{pmatrix}
\Bs-\Bb k\\ -\Bb k
\end{pmatrix} 
\ \text{and}\
\Phi(k;\delta_1,\delta_3)\begin{pmatrix}
C_1\\-\eta C_3\end{pmatrix} =\begin{pmatrix}
C_2 k^{\gamma_2}+
a_0\Bs-a_1\Bb k
\\C_2 \gamma_2 k^{\gamma_2}
-a_1\Bb k
\end{pmatrix}. 
\]
It follows that
\beq{C1C3-11}
\begin{pmatrix}
C_1\\ C_3\end{pmatrix} =\Phi^{-1} (k;\delta_1,\delta_3)
\begin{pmatrix}
\Bs-\Bb k\\ -\Bb k
\end{pmatrix},
\eeq
and
\beq{C1C3-21}
\begin{pmatrix}
C_1\\  C_3\end{pmatrix} =\begin{pmatrix}
1 & 0\\ 0 & -\eta^{-1}
\end{pmatrix}\Phi^{-1} (k;\delta_1,\delta_3)
\begin{pmatrix}
C_2 k^{\gamma_2}+
a_0\Bs -a_1\Bb k
\\C_2 \gamma_2 k^{\gamma_2}
-a_1\Bb k
\end{pmatrix}. 
\eeq
Eliminate $C_1$ and $C_3$ to obtain
\beq{C2k-1}
\Phi (k;\delta_1,\delta_3)\begin{pmatrix}
1 & 0\\ 0 & -\eta
\end{pmatrix}
\Phi^{-1} (k;\delta_1,\delta_3)
\begin{pmatrix}
\Bs-\Bb k\\ -\Bb k
\end{pmatrix}
=\begin{pmatrix}
C_2 k^{\gamma_2}+
a_0\Bs-a_1\Bb k
\\C_2 \gamma_2 k^{\gamma_2}
-a_1\Bb k
\end{pmatrix}. 
\eeq
Some simple calculations yield
\[\Phi (k;\delta_1,\delta_3)\begin{pmatrix}
1 & 0\\ 0 & -\eta
\end{pmatrix}
\Phi^{-1} (k;\delta_1,\delta_3)=
\frac{1}{\delta_3-\delta_1}
\begin{pmatrix}
\delta_3+\eta\delta_1 & -(1+\eta)\\
\delta_1\delta_3 (1+\eta) & -(\delta_1+\eta\delta_3)
\end{pmatrix}.
\]
We note that this matrix is independent of $k$. This reduces \rf{C2k-1} to
\[
\frac{1}{\delta_3-\delta_1}
\begin{pmatrix}
\delta_3+\eta\delta_1 & -(1+\eta)\\
\delta_1\delta_3 (1+\eta) & -(\delta_1+\eta\delta_3)
\end{pmatrix}
\begin{pmatrix}
\Bs-\Bb k\\ -\Bb k
\end{pmatrix}
=\begin{pmatrix}
C_2 k^{\gamma_2}+
a_0\Bs-a_1\Bb k
\\C_2 \gamma_2 k^{\gamma_2}
-a_1\Bb k 
\end{pmatrix}. 
\]
This leads to two equations:
\begin{align*}
\frac{(\delta_3+\eta\delta_1)(\Bs-\Bb k)+(1+\eta)\Bb k}{\delta_3-\delta_1}
&=C_2 k^{\gamma_2}+a_0\Bs-a_1\Bb k, \\
\frac{\delta_1\delta_3 (1+\eta)(\Bs-\Bb k)+(\delta_1+\eta\delta_3)\Bb k}{
\delta_3-\delta_1}
&=C_2 \gamma_2 k^{\gamma_2}-a_1\Bb k.
\end{align*}
Eliminating $C_2 k^{\gamma_2}$, we obtain an equation containing only $k$:
\begin{align*}
& \frac{[(\delta_3+\eta\delta_1)(\Bs-\Bb k)+(1+\eta)\Bb k]\gamma_2-
[\delta_1\delta_3 (1+\eta)(\Bs -\Bb k)+(\delta_1+\eta\delta_3)\Bb k]}{
\delta_3-\delta_1}\\
= & a_0\gamma_2\Bs-a_1(\gamma_2-1)\Bb  k.
\end{align*}

This leads to the solution 
\beq{k2}
k=\frac{\delta_1\delta_3(1+\eta)-(\delta_3+\eta\delta_1)\gamma_2-a_0(\delta_1-\delta_3)\gamma_2}
{(1+\eta)(\delta_1\delta_3 + \gamma_2)-(\delta_3+\eta\delta_1)\gamma_2
-(\delta_1+\eta\delta_3)-a_1(\gamma_2-1)(\delta_1-\delta_3)} \cdot \frac{\Bs}{\Bb}.
\eeq

Note that $k>0$ and this can be shown by proving that both the
numerator and denominator in \rf{k2}
are positive. Since $0<a_0<1$ and $0<a_1<1$, we have
\[
\begin{array}{l}
\hspace*{-0.5in}
  \delta_1\delta_3(1+\eta)-(\delta_3+\eta\delta_1)\gamma_2-a_0(\delta_1-\delta_3)\gamma_2\\
> \delta_1\delta_3(1+\eta)-(\delta_3+\eta\delta_1)\gamma_2-(\delta_1-\delta_3)\gamma_2\\
=(1+\eta)(-\delta_1)(\gamma_2-\delta_3)>0.
\end{array}
\]
Moreover, 
\[
\begin{array}{l}
\hspace*{-0.3in}
   (1+\eta)(\delta_1\delta_3 + \gamma_2)-(\delta_3+\eta\delta_1)\gamma_2
-(\delta_1+\eta\delta_3)-a_1(\gamma_2-1)(\delta_1-\delta_3)\\
> (1+\eta)(\delta_1\delta_3 + \gamma_2)-(\delta_3+\eta\delta_1)\gamma_2
-(\delta_1+\eta\delta_3)-(\gamma_2-1)(\delta_1-\delta_3)\\
= (1+\eta)(\delta_1\delta_3 + \gamma_2)-\gamma_2(\delta_3+\eta\delta_1+\delta_1-\delta_3)
-(\delta_1+\eta\delta_3-\delta_1+\delta_3)\\
= (1+\eta)(\delta_1\delta_3 + \gamma_2-\gamma_2\delta_1-\delta_3)\\
= (1+\eta)(\gamma_2-\delta_3)(1-\delta_1)>0.
\end{array}
\]
Therefore, $k>0$.

Next, we solve for the rest of parameters.
From \rf{C1C3-11}, we have
\beq{C1C3-1}
C_1= \frac{-\delta_3\Bs +(\delta_3-1)\Bb k}{(\delta_1-\delta_3)k^{\delta_1}},
\quad
\text{and}\quad 
C_3= \frac{\delta_1\Bs -(\delta_1-1)\Bb k}{(\delta_1-\delta_3)k^{\delta_3}}.
\eeq

Similarly, \rf{C1C3-21} yields
\beq{C1C3-2}
\begin{split}
C_1 &= \frac{(\gamma_2-\delta_3)C_2k^{\gamma_2}-a_0\delta_3 \Bs +a_1(\delta_3-1)\Bb k}{(\delta_1-\delta_3)k^{\delta_1}}, \\
C_3 &=\frac{(\gamma_2-\delta_1)C_2k^{\gamma_2}-a_0\delta_1 \Bs +a_1(\delta_1-1)\Bb k}{\eta (\delta_1-\delta_3)k^{\delta_3}}.
\end{split}
\eeq
Combine \rf{C1C3-1} and \rf{C1C3-2} to obtain
\begin{align*}
(\gamma_2-\delta_3)C_2k^{\gamma_2}+(1-a_1)(1-\delta_3)\Bb k &=(a_0-1)\delta_3\Bs, \\
(\gamma_2-\delta_1)C_2k^{\gamma_2}-(\eta+a_1)(1-\delta_1)\Bb k &=(a_0+\eta)\delta_1\Bs.
\end{align*}
Eliminating the term linear in $k$, we have
\beq{C2}
C_2=\frac{[(1-a_0)(\eta+a_1)(1-\delta_1)(-\delta_3)+(1-a_1)(\eta+a_0)(1-\delta_3)\delta_1]\Bs}{
(1+\eta)(\delta_1\delta_3 + \gamma_2)-\delta_1[(a_1+\eta)\gamma_2+(1-a_1)]-\delta_3[
(a_1+\eta)+\gamma_2(1-a_1)] k^{\gamma_2}}.
\eeq

Finally, we give a lemma needed in the proof a verification theorem to follow.
Its proof is technical and length. We provide it in the Appendix.

\begin{lem}\label{C1C2C3}\bdd
Under Assumption {\rm (A1)}, the constants $C_1$, $C_2$, and $C_3$ are positive.
\end{lem}

\section{A Verification Theorem}
In this section, we first show that the functions $w_0$ and $w_1$
are solutions of the HJB equations \rf{HJB-ode}.
Then, we provide a verification theorem.

\begin{thm}\label{HJBsolu}\bdd
  Assume {\rm (A1)}. Then, the following functions $w_0$ and $w_1$
  satisfy the HJB equations \rf{HJB-ode}:
\[
\begin{array}{l}
w_0(y)=\left\{\begin{array}{ll}
  C_2y^{\gamma_2}+a_0\Bs-a_1\Bb y&\mbox{ if }0<y<k,\\
  C_1y^{\delta_1}-\eta C_3y^{\delta_3}&\mbox{ if }y\geq k,\\
\end{array}\right.\\
w_1(y)=\left\{\begin{array}{ll}
  \Bs-\Bb y&\mbox{ if }0<y<k,\\
  C_1y^{\delta_1}+C_3y^{\delta_3}\hspace*{0.53in}&\mbox{ if }y\geq k.\\
  \end{array}\right.\\
\end{array}
\]
\end{thm}

\nd{\it Proof.}
It suffices to show the following variational inequalities hold:
\[ \begin{split}
(0,k):\quad & (\rho+\lambda_1-\L) w_1(y) \geq \lambda_1 w_0(y), \\
(k,\infty): \quad & w_1(y)\geq \Bs-\Bb y.
\end{split}
\]

Recall that on the interval $(0,k)$,
\[w_0(y)=C_2y^{\gamma_2}+a_0\Bs -a_1\Bb y\quad\text{and}\quad
w_1(y)=\Bs-\Bb y.\]
Then
\[ (\rho+\lambda_1-\L) w_1(y)=(\rho+\lambda_1-\mu_1)\Bs -(\rho+\lambda_1-\mu_2)\Bb y.\]
We let
\begin{align*}
\psi(y) &=(\rho+\lambda_1-\L) w_1(y)-\lambda_1 w_0(y)\\
       &=(\rho+\lambda_1-\mu_1)\Bs -(\rho+\lambda_1-\mu_2)\Bb y-\lambda_1(C_2y^{\gamma_2}+a_0\Bs -a_1\Bb y)
           \\
       &= [\rho+  (1-a_0)\lambda_1-\mu_1]\Bs-[\rho+(1-a_1)\lambda_1-\mu_2]\Bb y -\lambda_1C_2y^{\gamma_2}.
\end{align*}
We need to show that $\psi(y)\geq 0$  on the interval $(0,k)$. First, we note that
\[ \psi(0)=[\rho+  (1-a_0)\lambda_1-\mu_1]\Bs>0
\mbox{ and }
\psi'(0)=-[\rho+(1-a_1)\lambda_1-\mu_2]\Bb <0.
\]
We also have
\[\psi''(y)=-C_2\lambda_1\gamma_2(\gamma_2-1)y^{\gamma_2-2} <0\quad \text{since $\gamma_2>1$ and $C_2>0$}.\]
Hence $\psi'(y)$ is decreasing and $\psi'(y)<0$ on the interval $(0,k)$. 
It suffices to show that $\psi(k)\geq 0$ which implies $\psi(y)\geq 0$ for $0\leq y \leq k$. 
Introduce new functions $w^+_j$ and $w^-_j$ such that
\[w_{j}(y)=\begin{cases} w^{-}_j(y) & 0\leq y< k,\\
w^{+}_j(y) & y\geq k, \end{cases} \quad \text{for}\ j=0,1.
\]

Then, following from the smooth-fit conditions, we have, for $j=0,1$,
\[w^{-}_j(k)=w^{+}_j(k) \quad\text{and}\quad [w^{-}_j]'(k)=[w^{+}_j]'(k).\]
Moreover, $\psi(k)\geq 0$ is equivalent to
\[  (\rho+\lambda_1-\L)w^{-}_1(y)|_{y=k}\geq \lambda_1 w^{-}_0(k).\]
Note that 
\[
\quad (\rho+\lambda_1-\L)w^{+}_1(y)|_{y=k}=\lambda_1 w^{+}_0(y)|_{y=k}=\lambda_1 w^{-}_0(y)|_{y=k}.
\]

This reduces the proof of $\psi(k)\geq 0$ to
\[(\rho+\lambda_1-\L)w^{-}_1(y)|_{y=k} \geq (\rho+\lambda_1-\L)w^{+}_1(y)|_{y=k}.\]
Then we use 
\[w^{-}_j(k) := \lim_{y \uparrow k} w^{-}_j(y) =w^{+}_j(k) \quad\text{and}\quad [w^{-}_j]'(k)  := \lim_{y \uparrow k}[w^{-}_j]'(y) =[w^{+}_j]'(k),\]
for $j=0,1$,
to the above to get
\[-\sigma k^2 [w^{-}_1]''(y) |_{y=k} \geq -\sigma k^2 [w^{+}_1]''(y) |_{y=k},
\]
which is equivalent to
\[
[w^{-}_1]''(y) |_{y=k}  \leq [w^{+}_1]''(y) |_{y=k}.
\]
Recall \lemref{C1C2C3}.
The latter holds because
\[[w^{-}_1]''(y) |_{y=k} =0\quad \text{and}\quad
[w^{+}_1]''(y) |_{y=k}=C_1\delta_1(\delta_1-1)k^{\delta_1-2}+C_3\delta_3(\delta_3-1)k^{\delta_3-2}>0.\]
Therefore, $\psi(k)\geq0$ and hence $\psi(y)\geq0$ on $(0,k)$.

On the interval $(k,\infty)$ we need to show that $w_1(y)\geq \Bs-\Bb y$ with $
w_1(y)=C_1y^{\delta_1}+C_3 y^{\delta_3}$.
Let
$\phi(y)=C_1y^{\delta_1}+C_3 y^{\delta_3}-\Bs+\Bb y$. Then the smooth-fitting conditions implies $\phi(k)=\phi'(k)=0$. Moreover,
\begin{align*}
\phi'(y)&=C_1\delta_1y^{\delta_1-1}+C_3\delta_3 y^{\delta_3-1}+\Bb, \\
\phi''(y)&=C_1\delta_1(\delta_1-1)y^{\delta_1-2}+C_3\delta_3(\delta_3-1) y^{\delta_3-2}.
\end{align*}
Since $\delta_1<0$ and $\delta_3<0$, $\phi''(y)>0$ in the interval $[y,\infty)$.
This implies $\phi'(y)$ is increasing and $\phi'(k)=0$ implies $\phi'(y)>0$ for $y>k$.
Hence $\phi(y)$ is increasing, $\phi(k)=0$ implies $\phi(y)>0$ for $y>k$.
\hfill{$\Box$}

Next, we next provide a verification theorem.

\begin{thm}\label{VerificationThm}\bdd
Assume {\rm (A1)}.
Then, $v_\al(x_1,x_2)=x_1 w_\al(x_2/x_1)=V_\al(x_1,x_2)$, $\al=0,1$.
Let $D=\{(x_1,x_2,1):\ x_2>k x_1\}$.
Let $\tau^*=\inf\{t:\ (X^1_t,X^2_t,\al_t)\not\in D\}$.
Then $\tau^*$ is optimal.
\end{thm}

\nd{\it Proof.} 
The proof is similar to that of \cite[Theorem 2]{GuoZ}.
We only sketch the main steps for the sake of completeness.
First, for any admissible stopping time $\tau$,
following Dynkin's formula, we have
\[
v_\al(x_1,x_2)\geq E e^{-\rho\tau} v_{\al_\tau}(X^1_\tau,X^2_\tau)
\geq E e^{-\rho\tau} (\Bs X^1_\tau-\Bb X^2_\tau)=J(x_1,x_2,\al,\tau).
\]
So, $v_\al(x_1,x_2)\geq V_\al(x_1,x_2)$. The equality holds when $\tau=\tau^*$.
Hence, $v_\al(x_1,x_2)=J(x_1,x_2,\al,\tau^*)=V_\al(x_1,x_2)$.
\hfill{$\Box$}

\section{Asymptotics of $k=k(\la_0,\la_1)$}
In this section, we consider the asymptotic behavior of $k=k(\la_0,\la_1)$
as one of $\la_0$ and $\la_1$ goes to $\infty$ with the other fixed.
To facilitate the subsequent calculation, we
regroup  the terms in \rf{k2} for $k$ as follows:
\[k=\frac{(1+\eta)\delta_1\delta_3+\gamma_2[(a_0+\eta)(-\delta_1)+(1-a_0)(-\delta_3)]}
{(1+\eta)(\delta_1\delta_3 + \gamma_2)-\delta_1[(a_1+\eta)\gamma_2+(1-a_1)]-\delta_3[
(a_1+\eta)+\gamma_2(1-a_1)]}\cdot
\frac{\Bs}{\Bb}.\]

\subsection*{Asymptotics of $k$ as $\lambda_0\to\infty$}
We first consider the limit of $k=k(\la_0,\la_1)$ as $\la_0\to\infty$
with $\la_1$ fixed. In this case, the mean time for $\al_t$ spent at state 0 
is given by $(1/\la_0)$ which goes to 0. In view of this, the limit of $k$ should
correspond to the threshold of unconstrained pairs selling.

To validate this observation, we list all the terms in \rf{k2} that are
$\la_0$ depended:
\[\eta=\frac{\lambda_0}{\lambda_1},\
1- a_0=\frac{\rho-\mu_1}{\rho-\mu_1+\lambda_0},\
1-a_1=\frac{\rho-\mu_2}{\rho-\mu_2+\lambda_0},\
\gamma_2\approx \frac{\sqrt{\lambda_0}}{\sqrt{\sigma}},\
\text{and}\
\delta_3\approx -\frac{\sqrt{\lambda_0}}{\sqrt{\sigma}}.\]
Therefore, we have
\[
\lim_{\lambda_0\to\infty}
k=\lim_{\lambda_0\to\infty}\frac{-2\delta_1\lambda_0^{3/2}/(\lambda_1\sqrt{\sigma})+\text{lower order terms}}
{2(1-\delta_1)\lambda_0^{3/2}/ (\lambda_1\sqrt{\sigma})+ \text{lower order terms}} \cdot \frac{\Bs}{\Bb}
=\frac{-\delta_1}{1-\delta_1}\cdot \frac{\Bs}{\Bb}=:k_0.
\]
To see the connection with the selling rule without constraints.
We note that the associated HJB equation (unconstrained) has the form:
\[
\min\Big\{ \rho w(y)- \L w(y),\  w(y)-\Bs  +\Bb  y\Big\}=0.
\]
Repeating our previous smooth-fit calculation yields exactly $k=k_0$ obtained above.

\subsection*{Asymptotics of $k$ as $\lambda_1\to\infty$}
Similarly, we can consider the limit $\lambda_1\to\infty$
with fixed $\lambda_0$. 
Note that $\delta_3\approx -{\sqrt{\lambda_1}}/{\sqrt{\sigma}}$
and $\eta={\lambda_0}/{\lambda_1}$ are the only two terms depending on $\la_1$.
It follows that
\[
\lim_{\lambda_1\to\infty} k =
\frac{\delta_1-\gamma_2(1-a_0)}{\delta_1-[a_1+\gamma_2(1-a_1)]}\cdot \frac{\Bs}{\Bb}
=\frac{-\delta_1+\gamma_2(1-a_0)}{1-\delta_1+(\gamma_2-1)(1-a_1)} \cdot \frac{\Bs}{\Bb}=:k_1.
\]

It is not difficult to show $k_1>k_0$. (Actually, it is equivalent to the
inequality in \rf{ineq1} that is proved in the Appendix.) 
Intuitively, this makes sense because one has to make trading
easier (with larger $k$) when trading constraints are present. 

\section{Numerical Examples}

\begin{figure}[h]
     \centering
     \begin{subfigure}[b]{0.48\textwidth}
         \centering
         \includegraphics[width=\textwidth]{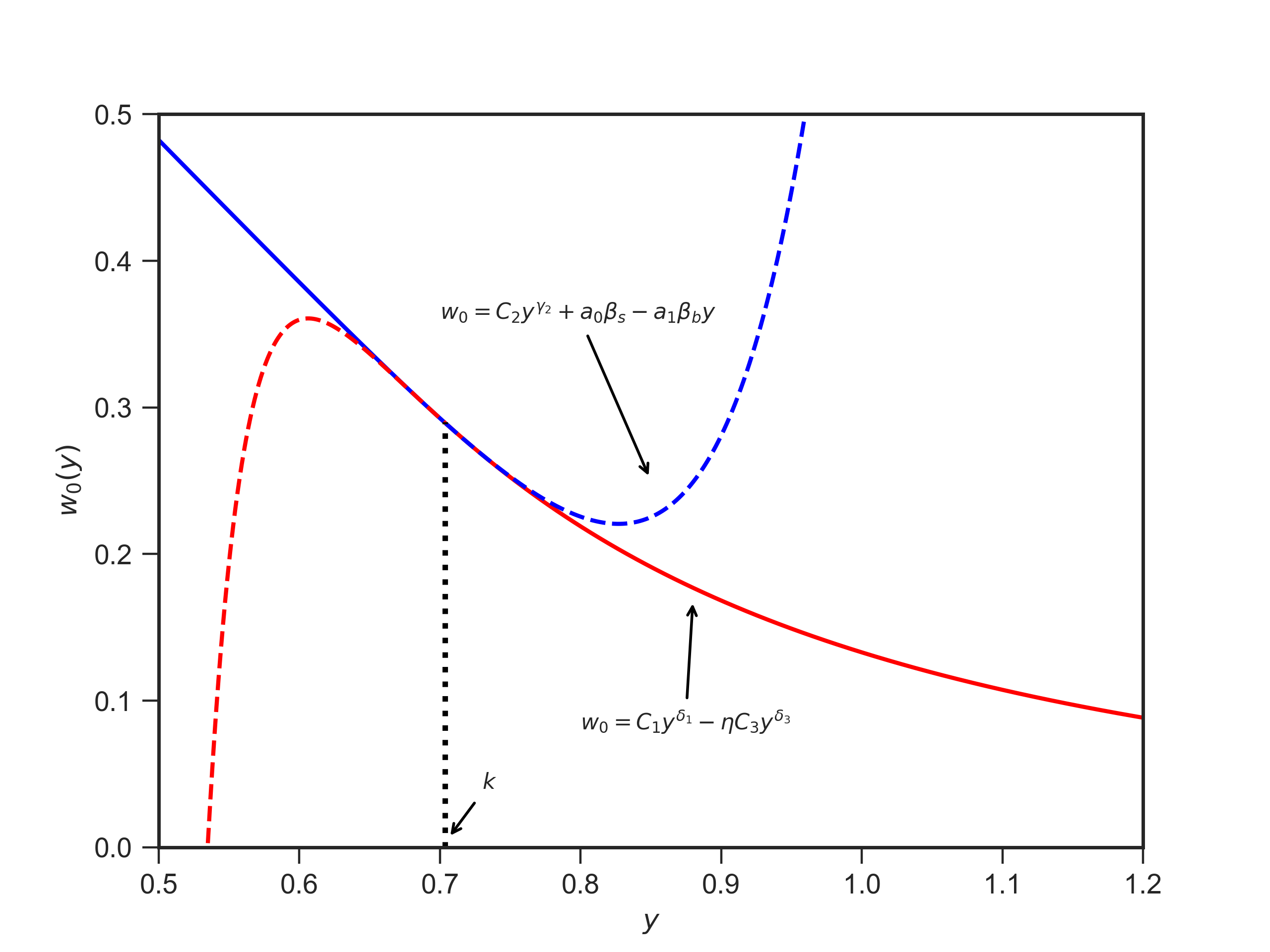}
     \end{subfigure}
     \hfill
     \begin{subfigure}[b]{0.48\textwidth}
         \centering
         \includegraphics[width=\textwidth]{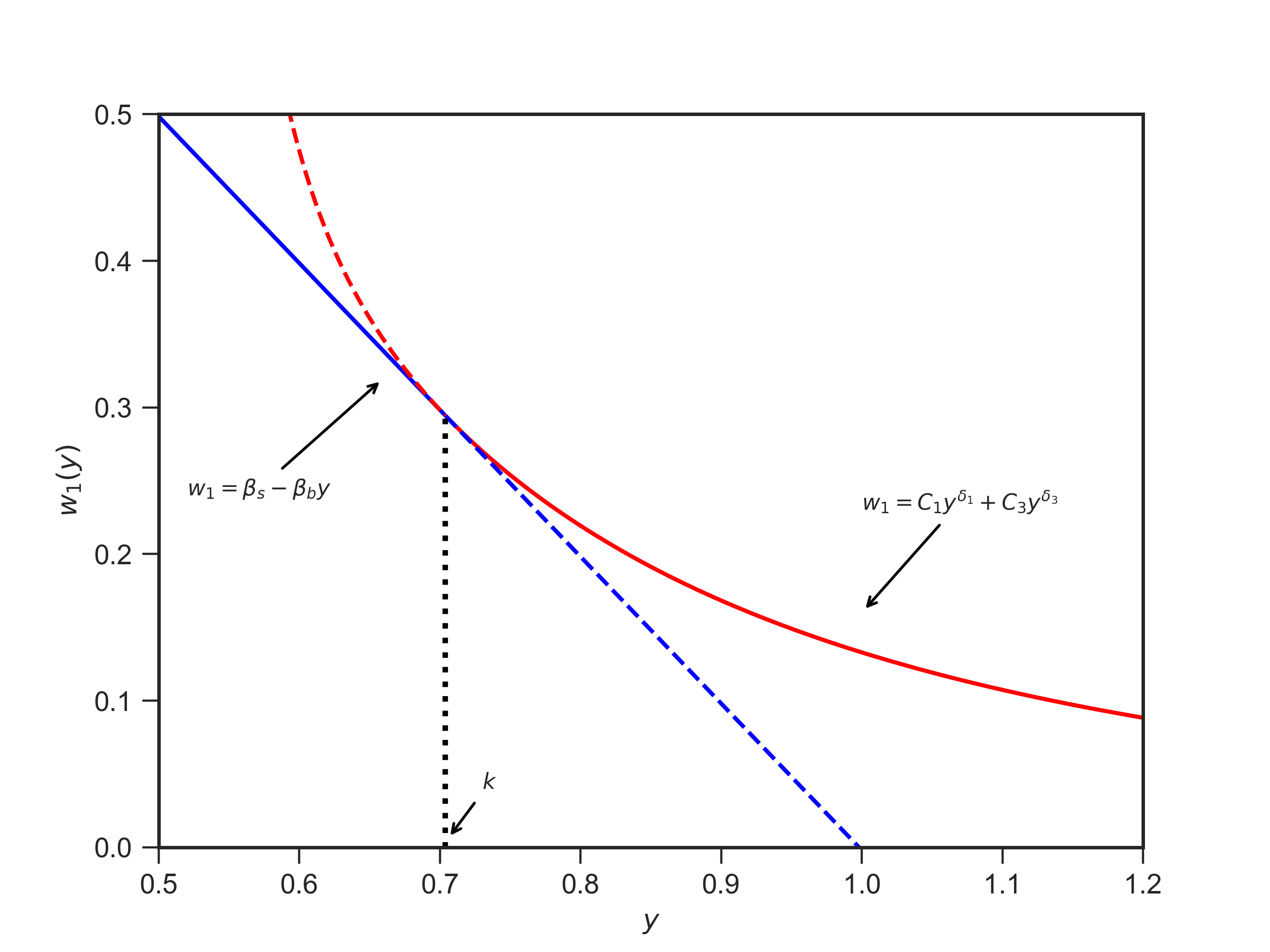}
     \end{subfigure}
        \caption{Functions $w_0$ and $w_1$}
        \label{w0w1-graph}
\end{figure}

\begin{figure}[!h]
     \centering
     \begin{subfigure}[b]{0.48\textwidth}
         \centering
         \includegraphics[width=\textwidth]{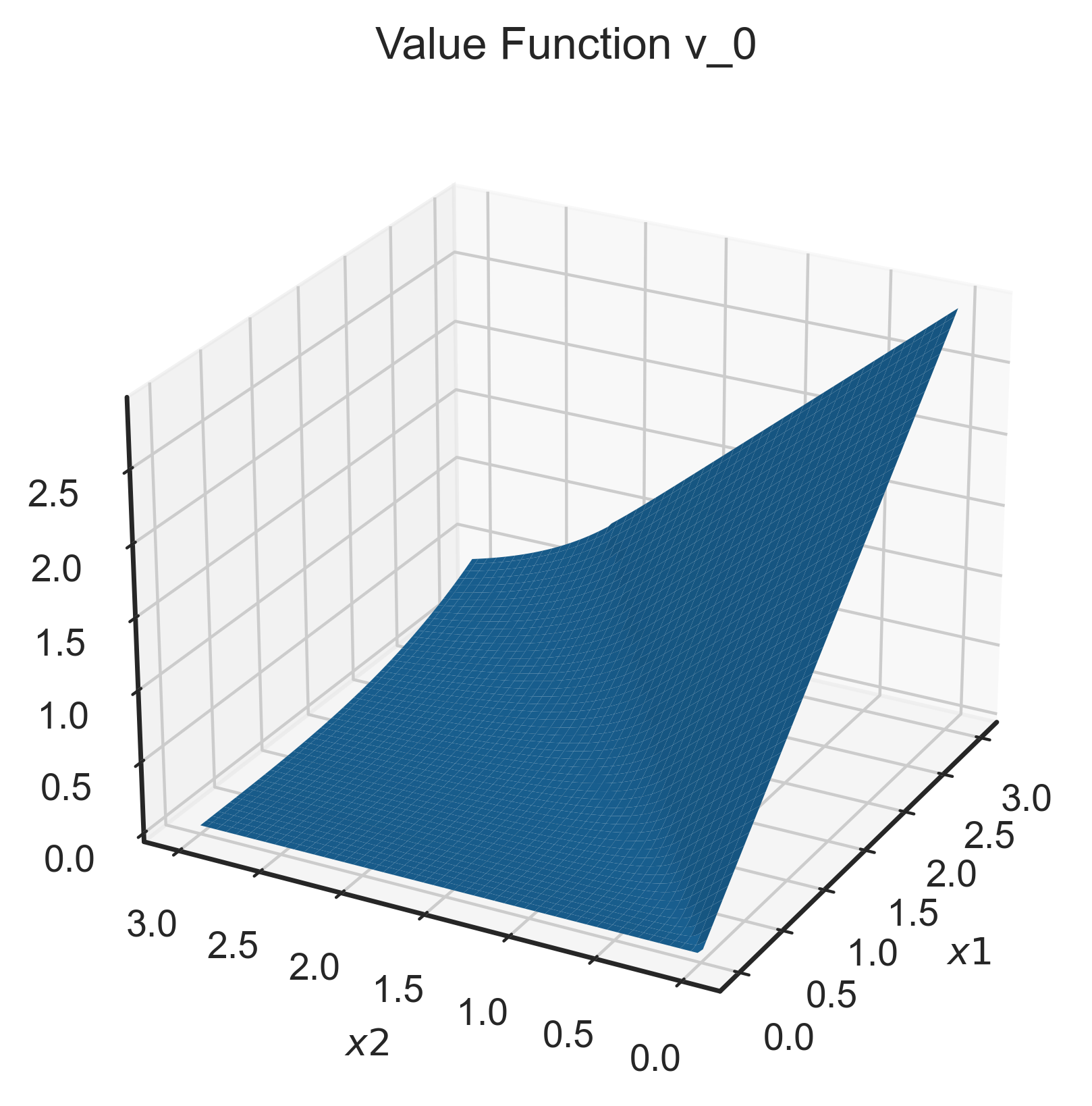}
     \end{subfigure}
     \hfill
     \begin{subfigure}[b]{0.48\textwidth}
         \centering
         \includegraphics[width=\textwidth]{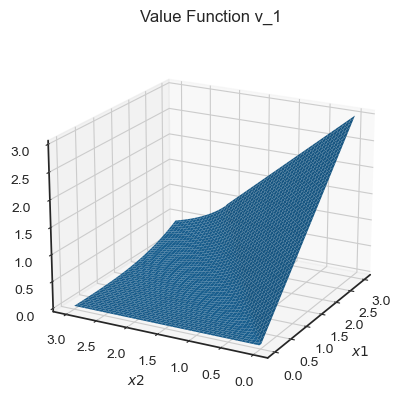}
     \end{subfigure}
        \caption{Functions $v_0$ and $v_1$}
        \label{v0v1}
\end{figure}

In this section, we consider the trading pair of Target Corp. (TGT) and Walmart Stores Inc. (WMT). The model is calibrated by using the daily closing prices from 1985-1999. Let $\bfS^1$=WMT and $\bfS^2$=TGT. Using the traditional least squares method, we have
$\mu_1=0.2459$, $\mu_2=0.2059$, $\sigma_{11}=0.2943$, $\sigma_{12}=0.0729$,
$\sigma_{21}=0.0729$, and $\sigma_{22}=0.3112$.
We take $\rho=0.5$, $\lambda_0=\lambda_1=10$, and $K=0.001$.
Using \rf{k2}, we obtain $k=0.7036$.
We plot the corresponding $w_0(y)$ and $w_1(y)$ in Figure~\ref{w0w1-graph} and
$v_i(x_1,x_2)=x_1 w_i(x_2/x_1), i=0, 1$ in Figure~\ref{v0v1}.

\begin{figure}[!h]
     \centering
     \begin{subfigure}[b]{0.48\textwidth}
         \centering
         \includegraphics[width=\textwidth]{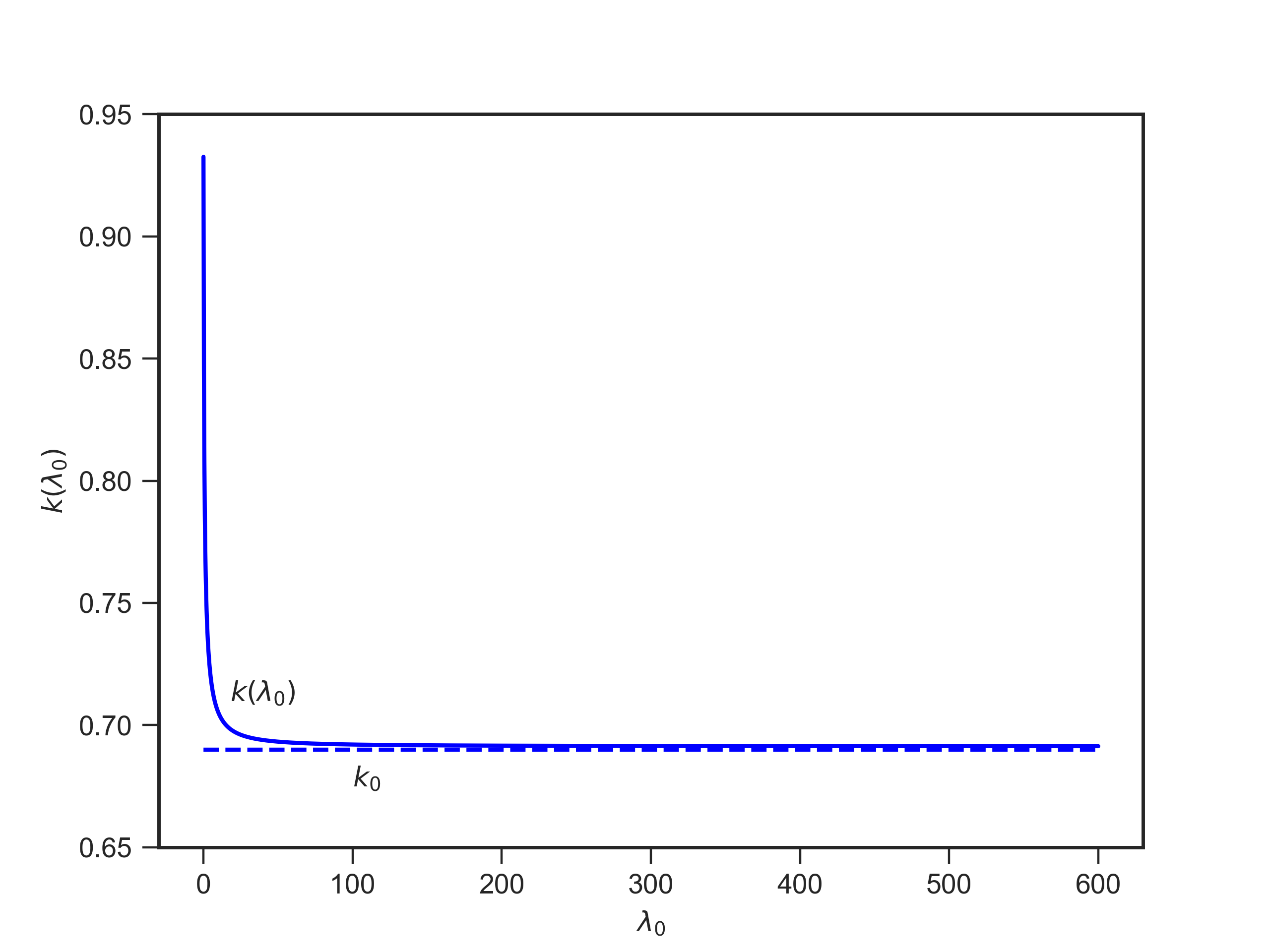}
     \end{subfigure}
     \hfill
     \begin{subfigure}[b]{0.48\textwidth}
         \centering
         \includegraphics[width=\textwidth]{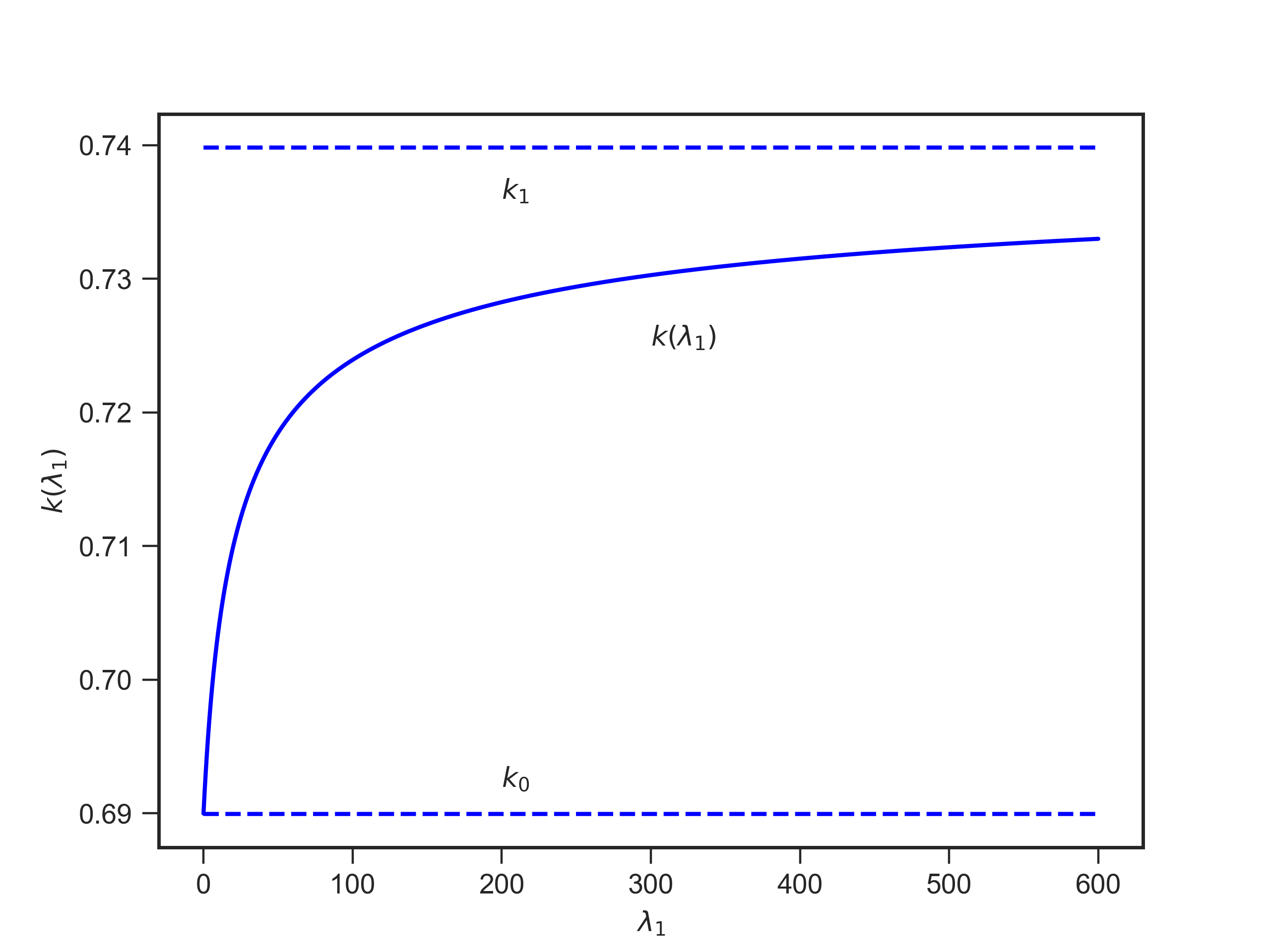}
     \end{subfigure}
        \caption{Asymptotics of $k(\lambda_0, 10)$ and $k(10, \lambda_1)$}
        \label{asy}
\end{figure}

\paragraph{Asymptotics of $k(\lambda_0, \lambda_1)$.}
Next, we examine the asymptotics of $k=k(\la_0,\la_1)$
as one of $\la_0$ and $\la_1$ goes to $\infty$ with the other fixed (at 10).
The corresponding graphs are plotted in Figure~\ref{asy} along with
$y=k_0$ and $y=k_1$.

\begin{figure}[!h]
	\centering
	\includegraphics[width=0.5\linewidth]{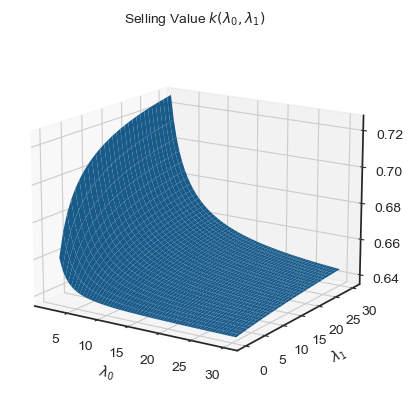}
	\caption{Dependence of $k$ on $\lambda_0, \lambda_1$}
	\label{k-value}
\end{figure}

In addition, we give the 2D graph of $k(\lambda_0, \lambda_1)$ in
Figure~\ref{k-value}. 

\paragraph{Dependence of $k$ on various parameters.}
In this section, we fix $\lambda_0=\lambda_1=10$. We vary one parameter at a time
and examine the dependence of $k$ on $\mu_i, \sigma_{ij}, i, j=1, 2, K, \rho$ respectively.

First, we vary $\mu_i, i=1, 2$. Recall that we close the position by selling $X^1$ and buying back $X^2$ on $[0, k]$ in the state $\alpha=1$. A larger $\mu_i$ implies greater growth potential in $X^i$. It can be seen in Table \ref{T1} that $k$ decrease in $\mu_1$ leading to fewer selling opportunities and $k$ increases in $\mu_2$ leading to more selling opportunities. It is because a larger $\mu_2$ and a smaller $\mu_1$ will encourage its early exit. 
{\footnotesize
\begin{table}[!h]
\caption{\bf $k$ with varying $\mu_1, \mu_2$}\label{T1}
\centering
\begin{tabular}{l| l l l l l}
\hline
 $\mu_1$ &0.1259 & 0.1659 &0.2059 &0.2459  & 0.2859 \\ [0.03in]\hline
 $k$ &0.7834 &0.7481 & 0.7036 & 0.6481  & 0.5798\\[0.01in]
  \hline \\[-0.08in]
  \hline
  $\mu_2$&0.1659 & 0.2059 &0.2459 &0.2859 & 0.3259\\[0.03in]  \hline
 $k$ &0.6332 &0.6688 & 0.7036 & 0.7367  & 0.7669 \\[0.01in]
   \hline
\end{tabular}
\end{table}
}

Next, we vary $\sigma_{11}$ and $\sigma_{22}$. Larger volatility
leads to a bigger room for the prices to move. This
is associated with a smaller selling zone and therefore smaller $k$.
{\footnotesize
\begin{table}[!h]
\caption{\bf $k$ with varying $\sigma_{11}, \sigma_{22}$}\label{T2}
\centering
\begin{tabular}{l| l l l l l}
\hline
 $\sigma_{11}$ &0.2312 & 0.2712 &0.3112 &0.3512  & 0.3912 \\ [0.03in]\hline
 $k$ &0.7516 &0.7286 & 0.7036 & 0.6777  & 0.6514\\[0.01in]
  \hline \\[-0.08in]
  \hline
  $\sigma_{22}$&0.2143 & 0.2543 &0.2943 &0.3343 & 0.3743\\[0.03in]  \hline
 $k$ &0.7469 &0.7265 & 0.7036 & 0.6794  & 0.6543 \\[0.01in]
   \hline
\end{tabular}
\end{table}
}

Then, we vary $\sigma_{12}$ $(=\sigma_{21})$. Increasing these parameters  makes the two stocks more related to each other which leads to a larger selling zone as shown in Table \ref{T3}. 
{\footnotesize
\begin{table}[!]
\caption{\bf $k$ with varying $\sigma_{12}=\sigma_{21}$}\label{T3}
\centering
\begin{tabular}{l| l l l l l}
\hline
 $\sigma_{12}=\sigma_{21}$ &-0.0129 & 0.0329 &0.0729 &0.1129  & 0.1529 \\ [0.03in]\hline
 $k$ &0.6060 &0.6561 & 0.7036 & 0.75477  & 0.8094\\ 
 \hline
\end{tabular}
\end{table}
}

Finally, we vary the discount rate $\rho$ and the transaction fee $K$. Larger $\rho$ encourages to sell early which translates to a bigger selling zone as can be seen in Table \ref{T4}. Additionally, larger $K$ sets up a higher barrier for selling which leads to a smaller selling zone. 
{\footnotesize
\begin{table}[h]
\caption{\bf $k$ with varying $\rho$ and $K $}\label{T4}
\centering
\begin{tabular}{l| l l l l l}
\hline
 $\rho$ & 0.3 & 0.4 & 0.5 & 0.6  & 0.7 \\ [0.03in]\hline
 $k$ &0.5590 &0.6541 & 0.7036 & 0.7358  & 0.7590\\[0.01in]
  \hline \\[-0.08in]
  \hline 
  $K$ &0.0001 & 0.0005 &0.001 &0.002 & 0.003\\[0.03in]  \hline
 $k$ &0.7049 &0.7043 & 0.7036 & 0.7022  & 0.7008 \\[0.01in]
   \hline
\end{tabular}
\end{table}
}

\section{Conclusion}

The focus of this paper is on pairs selling with limited opportunities, and its main goal is to derive an optimal policy in closed form, which is desirable for practical applications.

It would be intriguing to extend the results of this study to models that incorporate more practical considerations, such as large block selling, where intensive trading may affect trading windows.

Overall, this paper contributes to the understanding of pairs selling and provides insights into developing optimal policies that can be applied in real-world scenarios. There is potential for future research to expand on these findings and explore more complex models that better capture real-world dynamics.

\section*{Appendix}

In this appendix, we provide the proof of \lemref{C1C2C3}.

\nd{\it Proof of \lemref{C1C2C3}.}
We first work on $C_1$ and $C_3$.
In view of \rf{C1C3-1}, $C_1>0$ and $C_3>0$ are equivalent to
\beq{k-condition1} 
\frac{-\delta_1}{1-\delta_1}\cdot \frac{\beta_s}{\beta_b}< k<\frac{-\delta_3}{1-\delta_3}\cdot \frac{\beta_s}{\beta_b}.
\eeq

To simplify notation, let $b_0=1-a_0$ and $b_1=1-a_1$.
Note that $0<b_0<1$ and $0<b_1<1$. Using this notation, we rewrite $k$ as
\beq{k4}
k=\frac{(1+\eta)(-\delta_1)(\gamma_2-\delta_3)+b_0(\delta_1-\delta_3)\gamma_2}{(1+\eta)(\gamma_2-\delta_3)(1-\delta_1)+b_1(\gamma_2-1)(\delta_1-\delta_3)}\cdot \frac{\Bs}{\Bb}.
\eeq
The inequalities are equivalent to
\beq{ineq0}
\frac{-\delta_1}{1-\delta_1} <\frac{(1+\eta)(-\delta_1)(\gamma_2-\delta_3)+b_0(\delta_1-\delta_3)\gamma_2}{(1+\eta)(\gamma_2-\delta_3)(1-\delta_1)+b_1 (\gamma_2-1)(\delta_1-\delta_3)}
<\frac{-\delta_3}{1-\delta_3}.
\eeq

\subsection*{First inequality of \rf{ineq0}.}
The first inequality in \rf{ineq0} is equivalent to
\[b_1 (\gamma_2-1)(\delta_1-\delta_3)(-\delta_1)<b_0\gamma_2(\delta_1-\delta_3)(1-\delta_1)\
\Longleftrightarrow \ b_1 (\gamma_2-1)(-\delta_1)<b_0\gamma_2 (1-\delta_1).\]
The last inequality is equivalent to
\beq{ineq1} \frac{b_1}{b_0} <\frac{\gamma_2}{\gamma_2-1}\cdot (1-\frac{1}{\delta_1})
=\left(1+\frac{1}{\gamma_2-1}\right)\left(1-\frac{1}{\delta_1}\right).\eeq

Note that 
\[
  b_0=\frac{\rho-\mu_1}{\rho+\lambda_0-\mu_1}
  \mbox{ and } b_1 =\frac{\rho-\mu_2}{\rho+\lambda_0-\mu_2}.
  \]

  We consider two cases: Case I ($\mu_1\leq \mu_2$) and Case II ($\mu_1>\mu_2$).
  
{\bf Case I}: If $\mu_1\leq \mu_2$, then 
\[\frac{b_1}{b_0}=\frac{1+\frac{\lambda_0}{\rho-\mu_1}}{1+\frac{\lambda_0}{\rho-\mu_2}}\leq 1;\]
and the right hand side \rf{ineq1} is bigger than $1$ since $\gamma_2>1$ and $\delta_1<0$. So the first inequality in \rf{ineq0} follows.

{\bf Case II}: If $\mu_1>\mu_2$, then 
\[\frac{b_1}{b_0}=\frac{1+\frac{\lambda_0}{\rho-\mu_1}}{1+\frac{\lambda_0}{\rho-\mu_2}}=
\frac{(\rho+\lambda_0-\mu_1)(\rho-\mu_2)}{(\rho+\lambda_0-\mu_2)(\rho-\mu_1)}>1.\]
The previous simple argument no longer works. We need to elaborate
the value of 
$\left(1+\frac{1}{\gamma_2-1}\right)\left(1-\frac{1}{\delta_1}\right)$.
To this end, note that
\begin{align*}
\frac{\gamma_2}{\gamma_2-1} &=\frac{1+\frac{\mu_1-\mu_2}{\sigma}
+\sqrt{\left(1+\frac{\mu_1-\mu_2}{\sigma}\right)^2+\frac{4(\rho+\lambda_0-\mu_1)}{\sigma}}}{
-1+\frac{\mu_1-\mu_2}{\sigma}
+\sqrt{\left(1+\frac{\mu_1-\mu_2}{\sigma}\right)^2+\frac{4(\rho+\lambda_0-\mu_1)}{\sigma}}}\\
&=\frac{\rho+\lambda_0+\frac{\sigma+\sigma\sqrt{B_1}-\mu_1-\mu_2}{2}}{\rho+\lambda_0-\mu_2},
\end{align*}
where
\[B_1=\(1+\frac{\mu_1-\mu_2}{\sigma}\)^2+\frac{4(\rho+\lambda_0-\mu_1)}{\sigma}.\]

Next we compute
\begin{align*}
\frac{\delta_1-1}{\delta_1} &=\frac{
-1+\frac{\mu_1-\mu_2}{\sigma}-\sqrt{\(1+\frac{\mu_1-\mu_2}{\sigma}\)^2+\frac{4(\rho-\mu_1)}{\sigma}}}{
1+\frac{\mu_1-\mu_2}{\sigma}-\sqrt{\(1+\frac{\mu_1-\mu_2}{\sigma}\)^2+\frac{4(\rho-\mu_1)}{\sigma}} }\\
&=\frac{
1-\frac{\mu_1-\mu_2}{\sigma}+\sqrt{\(1+\frac{\mu_1-\mu_2}{\sigma}\)^2+\frac{4(\rho-\mu_1)}{\sigma}}}{
-1-\frac{\mu_1-\mu_2}{\sigma}+\sqrt{\(1+\frac{\mu_1-\mu_2}{\sigma}\)^2+\frac{4(\rho-\mu_1)}{\sigma}} }\\
&=\frac{\rho+\frac{\sigma+\sigma\sqrt{B_2}-\mu_1-\mu_2}{2}}{\rho-\mu_1},
\end{align*}
where
\[ B_2=\(1+\frac{\mu_1-\mu_2}{\sigma}\)^2+\frac{4(\rho-\mu_1)}{\sigma}.\]

Since $\mu_1>\mu_2$, we have 
\[\sqrt{B_1} >1+\frac{\mu_1-\mu_2}{\sigma} \quad\text{and}\quad
\sqrt{B_2} >1+\frac{\mu_1-\mu_2}{\sigma}.\]
This implies
\[\frac{\sigma+\sigma\sqrt{B_1}-\mu_1-\mu_2}{2}>\sigma-\mu_2, \quad
\frac{\sigma+\sigma\sqrt{B_2}-\mu_1-\mu_2}{2}>\sigma-\mu_2,\]
and
\begin{align*}
\left(1+\frac{1}{\gamma_2-1}\right)\left(1-\frac{1}{\delta_1}\right) & >
\frac{\rho+\lambda_0+\sigma-\mu_2}{\rho+\lambda_0-\mu_2} \cdot \frac{\rho+\sigma-\mu_2}{\rho-\mu_1}\\
&>\frac{(\rho+\lambda_0-\mu_1)(\rho-\mu_2)}{(\rho+\lambda_0-\mu_2)(\rho-\mu_1)}.
\end{align*}
This is exactly what we need to show. Here we have used
\[\rho+\lambda_0+\sigma-\mu_2>\rho+\lambda_0-\mu_1 \ \text{and}\ 
\rho+\sigma-\mu_2>\rho-\mu_2,\]
since $\mu_1>\mu_2$ and $\sigma>0$.

\subsection*{Second inequality of \rf{ineq0}.}
The second inequality in \rf{ineq0} is equivalent to
\begin{align*}
&(1+\eta)(-\delta_1)(\gamma_2-\delta_3)(1-\delta_3) +b_0\gamma_2(\delta_1-\delta_3)(1-\delta_3)\\
<&(1+\eta)(\gamma_2-\delta_3)(1-\delta_1)(-\delta_3)+b_1 (\gamma_2-1)(\delta_1-\delta_3)(-\delta_3).
\end{align*}
This is equivalent to
\[b_0\gamma_2(1-\delta_3)<(1+\eta)(\gamma_2-\delta_3)+b_1 (\gamma_2-1)(-\delta_3).\]
We will let $a=-\delta_3>0$ and $\gamma_2=1+\gamma$. Then, the above inequality is reduced to
\[b_0(1+\gamma)(1+a)<(1+\eta)(1+\gamma+a)+b_1\gamma a.\]
Since $1+\gamma+a=(1+a)(1+\gamma)-a\gamma$, the above inequality is equivalent to
\[(1+\eta-b_1)a\gamma<(1+\eta-b_0)(1+a)(1+\gamma)\quad \Longleftrightarrow\quad
\frac{1+\eta-b_1}{1+\eta-b_0}<\frac{(1+a)(1+\gamma)}{a\gamma}.\]
Then, we have
\begin{align*}
\frac{1+\eta-b_1}{1+\eta-b_0}&=\frac{1+\frac{\lambda_0}{\lambda_1}-
\frac{\rho-\mu_2}{\rho+\lambda_0-\mu_2}}{1+\frac{\lambda_0}{\lambda_1}-
\frac{\rho-\mu_1}{\rho+\lambda_0-\mu_1}}
=\frac{\frac{\lambda_0}{\lambda_1}+\frac{\lambda_0}{\rho+\lambda_0-\mu_2}}{
\frac{\lambda_0}{\lambda_1}+\frac{\lambda_0}{\rho+\lambda_0-\mu_1}}\\
&=\frac{1+\frac{\lambda_1}{\rho+\lambda_0-\mu_2}}{1+\frac{\lambda_1}{\rho+\lambda_0-\mu_1}}
=\frac{(\rho+\lambda_0+\lambda_1-\mu_2)(\rho+\lambda_0-\mu_1)}{(\rho+\lambda_0+\lambda_1-\mu_1)(\rho+\lambda_0-\mu_2)},
\end{align*}
and
\begin{align*}
\frac{(1+a)(1+\gamma)}{a\gamma} &=\left(1+\frac{1}{a}\right)\left(1+\frac{1}{\gamma}\right).
\end{align*}
Recall that
\begin{align*}
a=-\delta_3&=\frac{1}{2}\left[\sqrt{\(1+\frac{\mu_1-\mu_2}{\sigma}\)^2+\frac{4(\rho+\lambda_0+\lambda_1-\mu_1)}{\sigma}} - \left(1+\frac{\mu_1-\mu_2}{\sigma}\right)\right],\\
a+1&=\frac{1}{2}\left[\sqrt{\(1+\frac{\mu_1-\mu_2}{\sigma}\)^2+\frac{4(\rho+\lambda_0+\lambda_1-\mu_1)}{\sigma}} + \left(1-\frac{\mu_1-\mu_2}{\sigma}\right)\right].\\
\end{align*}
Simple calculation yields
\begin{align*}
\frac{1+a}{a} &=\frac{\sqrt{\(1+\frac{\mu_1-\mu_2}{\sigma}\)^2+\frac{4(\rho+\lambda_0+\lambda_1-\mu_1)}{\sigma}} + \left(1-\frac{\mu_1-\mu_2}{\sigma}\right)}{\sqrt{\(1+\frac{\mu_1-\mu_2}{\sigma}\)^2+\frac{4(\rho+\lambda_0+\lambda_1-\mu_1)}{\sigma}} - \left(1+\frac{\mu_1-\mu_2}{\sigma}\right)}\\
&=\frac{\left[\sqrt{\(1+\frac{\mu_1-\mu_2}{\sigma}\)^2+\frac{4(\rho+\lambda_0+\lambda_1-\mu_1)}{\sigma}} +1\right]^2-\left(\frac{\mu_1-\mu_2}{\sigma}\right)^2 }{\frac{4(\rho+\lambda_0+\lambda_1-\mu_1)}{\sigma}}\\
&=\frac{2+\frac{4(\rho+\lambda_0+\lambda_1-\mu_1)+2\mu_1-2\mu_2}{\sigma}+ 2\sqrt{B_3}}{\frac{4(\rho+\lambda_0+\lambda_1-\mu_1)}{\sigma}}\\
&=\frac{\rho+\lambda_0+\lambda_1-\frac{\mu_1+\mu_2}{2}+ \frac{1+\sqrt{B_3}}{2}\sigma}{\rho+\lambda_0+\lambda_1-\mu_1},
\end{align*}
where
\[B_3=\(1+\frac{\mu_1-\mu_2}{\sigma}\)^2+\frac{4(\rho+\lambda_0+\lambda_1-\mu_1)}{\sigma}.\]

Recall
$B_1=\(1+\frac{\mu_1-\mu_2}{\sigma}\)^2+\frac{4(\rho+\lambda_0-\mu_1)}{\sigma}$.
We have
\begin{align*}
\frac{\gamma+1}{\gamma} &=\frac{\frac{\mu_1-\mu_2}{\sigma}+1
+\sqrt{\left(\frac{\mu_1-\mu_2}{\sigma}+1\right)^2+\frac{4(\rho+\lambda_0-\mu_1)}{\sigma}}
}{
\frac{\mu_1-\mu_2}{\sigma}-1
+\sqrt{\left(\frac{\mu_1-\mu_2}{\sigma}+1\right)^2+\frac{4(\rho+\lambda_0-\mu_1)}{\sigma}}}\\
&=\frac{\rho+\lambda_0-\frac{\mu_1+\mu_2}{2}+ \frac{1+\sqrt{B_1}}{2}\sigma}{\rho+\lambda_0-\mu_2}.
\end{align*}
Note that
\[\sqrt{B_3}>\left|1+\frac{\mu_1-\mu_2}{\sigma}\right| \quad \text{and}\quad 
\sqrt{B_1}>\left|1+\frac{\mu_1-\mu_2}{\sigma}\right|.\]
This implies
\begin{align*}
\frac{(1+\sqrt{B_3})\sigma}{2}-\frac{\mu_1+\mu_2}{2} 
&> \frac{\sigma+|\sigma+\mu_1-\mu_2|-\mu_1-\mu_2}{2}\\
&=\begin{cases}
\sigma-\mu_2 &\sigma+\mu_1>\mu_2,\\
-\mu_1 &\sigma+\mu_1 \leq \mu_2.
\end{cases}
\end{align*}
and similarly
\[\frac{(1+\sqrt{B_1})\sigma}{2}-\frac{\mu_1+\mu_2}{2} >
\begin{cases}
\sigma-\mu_2 &\sigma+\mu_1>\mu_2,\\
-\mu_1 &\sigma+\mu_1 \leq \mu_2.
\end{cases}\]

This implies that if $\sigma+\mu_1>\mu_2$ we have
\begin{align*}
\frac{(1+a)(1+\gamma)}{a\gamma} &>\frac{\rho+\lambda_0+\lambda_1-\mu_2+ 
\sigma}{\rho+\lambda_0+\lambda_1-\mu_1} \cdot
\frac{\rho+\lambda_0-\mu_2+ \sigma}{\rho+\lambda_0-\mu_2}\\
&>\frac{(\rho+\lambda_0+\lambda_1-\mu_2)(\rho+\lambda_0-\mu_1)}{(\rho+\lambda_0+\lambda_1-\mu_1)(
(\rho+\lambda_0-\mu_2)}\\
&=\frac{1+\eta-b_1}{1+\eta-b_0}.
\end{align*}

Here we have used $\sigma+\mu_1-\mu_2>0$ hence $\sigma-\mu_2>-\mu_1$ and 
\[\rho+\lambda_0-\mu_2+\sigma>\rho+\lambda_0-\mu_1.\]

If $\sigma+\mu_1\leq \mu_2$, then $\mu_1<\sigma+\mu_1\leq \mu_2$ and
\[\rho+\lambda_0+\lambda_1-\mu_1>\rho+\lambda_0+\lambda_1-\mu_2\]
and
\begin{align*}
\frac{(1+a)(1+\gamma)}{a\gamma} &>\frac{\rho+\lambda_0+\lambda_1-\mu_1}{\rho+\lambda_0+\lambda_1-\mu_1} \cdot
\frac{\rho+\lambda_0-\mu_1}{\rho+\lambda_0-\mu_2}\\
&>\frac{(\rho+\lambda_0+\lambda_1-\mu_2)(\rho+\lambda_0-\mu_1)}{(\rho+\lambda_0+\lambda_1-\mu_1)(
(\rho+\lambda_0-\mu_2)}\\
&=\frac{1+\eta-b_1}{1+\eta-b_0}.
\end{align*}
This exactly what we need to prove for the second inequality in \rf{ineq0}.

\subsection*{Proof of $C_2>0$}
Recall $C_2$ given in \rf{C2}.
In that formula, the denominator is positive since $\delta_3<\delta_1<0$, $\gamma_2>1$,
and $0<a_0,\ a_1<1$. $C_2$ is positive if 
\[ (1-a_0)(\eta+a_1)(1-\delta_1)(-\delta_3)>(1-a_1)(\eta+a_0)(1-\delta_3)(-\delta_1).\]
Since $\delta_1<0$ and $\delta_3<0$, this is equivalent to
\beq{C2-1}
\frac{\eta+a_1}{1-a_1}\cdot \frac{-\delta_1}{1-\delta_1} >\frac{\eta+a_0}{1-a_0}
\cdot\frac{1-\delta_3}{-\delta_3}.
\eeq
We need to compute both sides and compare them.
Note that
\begin{align*}
\frac{\eta+a_1}{1-a_1}&=\frac{\frac{\lambda_0}{\lambda_1}+\frac{\lambda_0}{\rho+\lambda_0-\mu_2}}
{\frac{\rho-\mu_2}{\rho+\lambda_0-\mu_2}}=\frac{\lambda_0}{\lambda_1} \cdot
\frac{\rho+\lambda_0+\lambda_1-\mu_2}{\rho-\mu_2},\\
\frac{\eta+a_0}{1-a_0}&=\frac{\frac{\lambda_0}{\lambda_1}+\frac{\lambda_0}{\rho+\lambda_0-\mu_1}}
{\frac{\rho-\mu_1}{\rho+\lambda_0-\mu_1}}=\frac{\lambda_0}{\lambda_1} \cdot
\frac{\rho+\lambda_0+\lambda_1-\mu_1}{\rho-\mu_1}.
\end{align*}
Then \rf{C2-1} is reduced to
\beq{C2-3}
\frac{\rho+\lambda_0+\lambda_1-\mu_2}{\rho-\mu_2}\cdot \frac{1-\delta_1}{-\delta_1}>
\frac{\rho+\lambda_0+\lambda_1-\mu_1}{\rho-\mu_1}
\cdot\frac{1-\delta_3}{-\delta_3}. 
\eeq

Recall that
\[
\frac{1-\delta_1}{-\delta_1}=\frac{\rho-\frac{\mu_1+\mu_2}{2}+ \frac{1+\sqrt{B_2}}{2}\sigma}{\rho-\mu_1},
\mbox{ and }
\frac{1-\delta_3}{-\delta_3}
=\frac{\rho+\lambda_0+\lambda_1-\frac{\mu_1+\mu_2}{2}+ \frac{1+\sqrt{B_3}}{2}\sigma}{\rho+\lambda_0+\lambda_1-\mu_1}.
\]
Then \rf{C2-3} is reduced to
\[ \frac{\rho+\lambda_0+\lambda_1-\mu_2}{\rho-\mu_2} \left( \rho-\frac{\mu_1+\mu_2}{2}+ \frac{1+\sqrt{B_2}}{2}\sigma\right)>\rho+\lambda_0+\lambda_1-\frac{\mu_1+\mu_2}{2}+ \frac{1+\sqrt{B_3}}{2}\sigma.\]
This is equivalent to
\begin{align*}
&(\rho+\lambda_0+\lambda_1-\mu_2)\( \rho-\frac{\mu_1+\mu_2-\sigma}{2}\)-
\(\rho+\lambda_0+\lambda_1-\frac{\mu_1+\mu_2-\sigma}{2}\)(\rho-\mu_2)\\
>& (\rho-\mu_2)\frac{\sqrt{B_3}}{2}\sigma - (\rho+\lambda_0+\lambda_1-\mu_2)\frac{\sqrt{B_2}}{2}\sigma\\
=&\frac{\sigma}{2} (\rho-\mu_2)(\sqrt{B_3}-\sqrt{B_2})-(\lambda_0+\lambda_1)\frac{\sqrt{B_2}}{2}\sigma.
\end{align*}
We can simplify the above inequality and it is equivalent to
\beq{C2-4}
(\lambda_0+\lambda_1)(1+\frac{\mu_2-\mu_1}{\sigma}+\sqrt{B_2})>(\rho-\mu_2)(\sqrt{B_3}-\sqrt{B_2}).
\eeq

Note that
\[\sqrt{B_3}-\sqrt{B_2}=\frac{B_3-B_2}{\sqrt{B_3}+\sqrt{B_2}}=\frac{4(\lambda_0+\lambda_1)}{\sigma} \cdot
\frac{1}{\sqrt{B_3}+\sqrt{B_2}}.\]
We can reduce \rf{C2-4} to
\[\left(1+\frac{\mu_2-\mu_1}{\sigma}+\sqrt{B_2}\right)(\sqrt{B_3}+\sqrt{B_2})>\frac{4(\rho-\mu_2)}{\sigma}.\]
This is equivalent to
\[\left(1+\frac{\mu_2-\mu_1}{\sigma}+\sqrt{B_2}\right)\sqrt{B_3}+\left(1+\frac{\mu_2-\mu_1}{\sigma}\right)\sqrt{B_2} >\frac{4(\rho-\mu_2)}{\sigma}-B_2.\]
Then we note that
\begin{align*}
&\frac{4(\rho-\mu_2)}{\sigma}-B_2 =\frac{4(\rho-\mu_2)}{\sigma}-\(1+\frac{\mu_1-\mu_2}{\sigma}\)^2-\frac{4(\rho-\mu_1)}{\sigma}\\
=& \frac{4(\mu_1-\mu_2)}{\sigma} -\(1+\frac{\mu_1-\mu_2}{\sigma}\)^2=- \(1-\frac{\mu_1-\mu_2}{\sigma}\)^2=-\(1+\frac{\mu_2-\mu_1}{\sigma}\)^2.
\end{align*}
The inequality is equivalent to
\[\left(1+\frac{\mu_2-\mu_1}{\sigma}+\sqrt{B_2}\right)\sqrt{B_3}+\left(1+\frac{\mu_2-\mu_1}{\sigma}\right)\sqrt{B_2} >-\(1+\frac{\mu_2-\mu_1}{\sigma}\)^2.\]
Moving the term $-\left(1+\frac{\mu_2-\mu_1}{\sigma}\right)^2$ to the left-hand side and we obtain
\beq{C2-5}
\left(1+\frac{\mu_2-\mu_1}{\sigma}+\sqrt{B_2}\right)\left(1+\frac{\mu_2-\mu_1}{\sigma}+\sqrt{B_3}\right) >0.\eeq
Then we apply
\[\sqrt{B_2} >|1+\frac{\mu_1-\mu_2}{\sigma}|\quad \text{and}\quad
\sqrt{B_3} >|1+\frac{\mu_1-\mu_2}{\sigma}|\]
to obtain
\[1+\frac{\mu_2-\mu_1}{\sigma}+ \sqrt{B_2} >
\begin{cases} 2 &\text{if $\sigma >\mu_2-\mu_1$},\\
\frac{2(\mu_2-\mu_1)}{\sigma}\geq 2 &\text{if $\sigma\leq \mu_2-\mu_1$}.
\end{cases}
\]
Similarly, $1+\frac{\mu_2-\mu_1}{\sigma}+ \sqrt{B_3} >2$.
So \rf{C2-5} follows. Therefore, $C_2>0$.
\hfill{$\Box$}

\section*{Acknowledgments}

This work is supported jointly by the Australian Research Council Discovery Project DP200101550, the Natural Science Foundation of China 11831010 and 61961160732, the Natural Science Foundation of Shandong Province ZR2019ZD42 and the Taishan Scholars Climbing Program of Shandong TSPD20210302.

\end{document}